\documentclass[english,numbers,authoryear]{elsarticle}
\usepackage[T1]{fontenc}
\usepackage[latin9]{inputenc}
\usepackage{array}
\usepackage{booktabs}
\usepackage{graphicx}
\usepackage{esint}

\makeatletter

\newcommand{\apj}{ApJ}

\newcommand{\apjl}{ApJL}
\newcommand{\jgr}{J. Geophys. Res.}
\newcommand{\aap}{A~\&~A}
\newcommand{\apss}{Astrophys.~\&~ Space~Sci.}

\newcommand{\mnras}{MNRAS}
\newcommand{\solphys}{Sol.Phys.}

\newcommand{\ssr}{Space Sci. Rev.}

\newcommand{\prl}{Phys.Rev. Lett.}
\newcommand{\grl}{GeoPphys.Rev. Lett.}
\newcommand{\nat}{Nature}
\newcommand{\nar}{New Astr. Rev.}

\begin{document}
\begin{frontmatter}

\let\WriteBookmarks\relax
\def\floatpagepagefraction{1}
\def\textpagefraction{.001}
%\shorttitle{Zonal harmonics  for solar cycle forecast. }
%\shortauthors{VN Obridko et~al.}

\title{Zonal harmonics of solar magnetic field for solar cycle forecast }                      

\author[1,5]{VN Obridko}
\ead{obridko@izmiran.ru}
\address[1]{Izmiran, Kaluzhskoe Sh., 4, Troitsk, Moscow, 108840, Russia}
\author[1,3,4]{DD Sokoloff}
\ead{sokoloff.dd3@gmail.com}
\author[2]{VV Pipin}
\address[2]{Institute of Solar-Terrestrial Physics, Russian Academy of
Sciences, Irkutsk, 664033, Russia}
\ead{pip@iszf.irk.ru}
\author[1,3,4]{AS Shibalova}
\ead{as.shibalova@physics.msu.ru}
\address[3]{Department of Physics, Moscow State University, 19991, Russia}

\address[4]{Moscow Center of Fundamental and Applied Mathematics, Moscow,
119991, Russia}
\address[5]{Central Astronomical Observatory of the Russian Academy of Sciences at Pulkovo, St. Petersburg}
\author[1]{IM Livshits}

\cortext[c]{Corresponding author}

\begin{abstract}
According to the scheme of action of the solar dynamo, the poloidal magnetic field
can be considered a source of production of the toroidal magnetic field by the solar differential rotation. From the polar magnetic field proxies, it is natural to expect that solar Cycle 25 will be weak as recorded in sunspot data.
We suggest that there are parameters of the  zonal harmonics of the solar surface magnetic field, such as the magnitude of the $l=3$ harmonic or the effective multipole index, that can be used as a reasonable addition to the polar magnetic field proxies.
We discuss also some specific features of solar activity indices in Cycles 23 and 24.
\end{abstract}
\end{frontmatter}

\section{Introduction}

The problem of forecasting solar activity is a long-lasting one.
Actually, this problem occurred as soon as the solar cycle was discovered, but we are still far from its definite solution. Before each sunspot maximum, forecasts of the cycle amplitude appear, but the predicted values are in quite a wide range \citep{O95, LR98}. The past two Cycles, 23 and 24, were no exception.

{
Strictly speaking, all prediction methods can be more or less reasonably referred to one of the two groups. The first group can be conventionally called mathematical or, more precisely, statistical. In this case, the forecaster chooses a solar activity index (usually, the sunspot number - SSN) and extends the chosen series to one or several cycles ahead by various statistical methods. Here, the basic data are the sunspot number series, and no additional physical considerations are taken into account. There are lots of such publications. Their authors sometimes use quite complicated statistical methods, such as neural network or machine learning (e.g., the relatively recent papers by \citealp{Quass2007,Attia2013,Dani2019} and \citealp{Covas2019}) and low-dimensional solar attractor representation (\citealp{Kurths1990}; reviews by \citealp{C98, Hath09,Kane07}).}

{
More sophicticated statistical methods use some statistical relationships between the particular points and phases of the solar cycle: the Waldmeier effect - the relation between the activity rise rate in the early phase of a solar cycle till its maximum and the amplitude of the maximum, the rule of Gnevyshev-Ohl - the relation between the total number of sunspots during an even numbered cycle and during the following odd-numbered one (\citealp{GnOl48,Ohl66}, see  \citealp{O95} for review) or its better known and widely used modification - the relation between the amplitudes of an even numbered and the following odd numbered cycles \citep{K50}, the concept of the "principal phases of a cycle" \citep{Kuklinetal90}, etc. These forecasts sometimes can provide  a physical explanation of the observed statistical relationships.}

{A fundamentally different group of forecasts is based on understanding the processes of the magnetic field generation on the Sun. According to the solar dynamo theory developed by Parker (1955), the sunspot cycle is an oscillation between the toroidal and poloidal components.  Two processes are involved in the solar dynamo: generation of the toroidal field from the poloidal one and regeneration of the poloidal field of a new cycle with the opposite polarity from this toroidal field. So, if we know the solar polar field, we can predict the following sunspot maximum. Thus, prediction of the height of the following cycle is reduced to calculating or directly measuring the polar field during the minimum of the SSN cycle. Unfortunately, the existing studies in this field are very contradictory (e.g.,
\cite{DG06,  Cetal07, Jetal07, Detal08,Saf18}). The problem is associated with the need to take into account the stochastic component, which drives the dynamo out of the deterministic regime, and with uncertainties in the input parameters (\citealp{BT07, KN12, Pipin2012}), while the direct measurements of the polar magnetic field are not accurate enough for prediction (see, e.g.,  \citealp{Pevtsov2021}). }  

{
A special group is formed by "precursor methods", in which various parameters observed before the solar cycle maximum ("precursors") are isolated, and their correlations with spot-formation indices are used to forecast the amplitude of the following cycle maximum. Such a method was first proposed by Ohl, who noticed a correlation between the cycle amplitude and the minimum level of geomagnetic activity at the beginning of the cycle \citep{Ohl66} or the level of geomagnetic activity in the late declining phase of the previous cycle \citep{OO79}. The characteristics of the large-scale magnetic field (e.g., see \citealp{Metal01, Metal02, OS92} and references therein) as well as the meridional-circulation asymmetry \citep{OS09} can also serve as precursors, as will be shown below. }

{
A detailed classification of the prediction methods is given by \cite{P08}, who separates the climatology, precursor, theoretical (dynamo model), spectral, neural network, and stock market prediction methods. All prediction methods can be generically divided into precursor and statistical (including the majority
of the above classifications) techniques or their combinations \citep{Hetal99,Hath09}. The fact that the prediction of solar cycle did not improve with adding
more data (a new solar cycle) suggests that all present-day methods are unable to give
reliable prognoses (see for example modern extensive review of \cite{N21}. }

\section{The problem of Cycle 25}
{
There is strong evidence that the intensity of the solar polar field near the  time of minimum of a sunspot cycle determines the strength of the following activity cycle. If we predict the poloidal component (polar field) at the minimum of a cycle, then we can use it to forecast the toroidal component, which generates the sunspot cycle. Some predictions of the polar filed at the minimum of Cycle 24 suggest that Cycle 25 will be weaker than Cycle 24 \citep{HU16, Cetal16, Ietal17}. It is interesting that the same conclusion can be drawn on the base of the helioseismic inversion of the zonal flow variations near the base of the convection zone \citep{Kosovichev2019} and the results of the nonlinear dynamo model of the extended solar cycle by \cite{Pipin2020}.  
}

{
Starting from Cycle 22 the height of the cycles began to decrease violating the Gnevyshev-Ohl rule. \cite{D03} shows that the ratio of heights of the even and odd cycles changed with time and could be violated not only in the pair of Cycles 22-23, but also in the pair of Cycles 24-25. The height of Cycle 24 was predicted equal to $87.5 \pm 23.5$, \cite{OS09} who actually continues the study by \citealp{Metal01, Metal02}.
%deal with anomalies in the solar magnetic field and the meridional-circulation asymmetry. 
They interpreted the data as indication to a possible advent of a Maunder-type minimum or at least a sequence of a few low cycles. The process of transition was analyzed in detail by \cite{Jetal16}, who arrived at a conclusion that the first signatures of transition to a Grand Minimum had been noticed as early as in the 1960-ies. In this work and in a number of others  \citep{Georgetal15,Ietal17, Metal21} the situation on the Sun during the past cycles is characterized as transition to a period of relatively low solar activity. The possibility of the Maunder-type minimum can not be rule out but is less likely.
}

{
Thus, we can expect one or two cycles of moderate or low activity at the beginning of the XXI century. This resembles the Dalton minimum at the beginning of the XIX century. However, a more profound decline of the type of Maunder minimum could not be ruled out.
A decrease in solar activity can be also traced by changes in the solar-related geophysical parameters. This was noted in many papers, including some of those cited above. Various geophysical indices and their prognostic value were analyzed in \cite{Oetal13, Ketal13, Georgetal15, Ketal15, Ketal18,Georgetal18}. }

{
A decrease in solar activity can be also traced by changes in the solar-related geophysical parameters. This was noted in many papers, including some of those cited above. Various geophysical indices and their prognostic value were analyzed in \citep{Oetal13,Ketal13,Ketal18,Georgetal15,Georgetal18}. }

Recently \cite{N21} has studied  the situation with the forecast of Cycle 25 in detail. He analyzed 77 predictions. Most of them (66 papers) predict that Cycle 25 will be higher than Cycle 24. The sunspot number (SSN) at the maximum of Cycle 24 was 113.3 in the revised scale, the average value for all forecasts is 136.2. 
Nandy divides all types of forecast into 6 categories: Physical Model Based Forecasts, Precursor Technique Based Forecasts, Non-linear Model Based Forecasts, Statistical Forecasts, Spectral Methods Based Forecasts, Machine Learning and Neural Network Based Forecasts. The best internal convergence is found in the first type of forecasts. They indicate that Cycle 25 will virtually coincide with Cycle 24, differing slightly from the latter. The average value according to the forecasts of this group is 113; i.e., it fully coincides with Cycle 24.

Solar activity is a set of complex interrelated processes that manifest themselves in various physical information channels. These processes are governed by intricate relations between different agents, of which the magnetic field is undoubtedly the main one. The variety of processes involved in the solar activity is due to the fact that the magnetic field itself is multicomponent and is generated by processes depending on a large number of sometimes poorly known parameters. The components of the magnetic field vary depending on spatial characteristics and time. An analysis of the structure, relationship, and interaction of these components is necessary both to understand the generation of the magnetic field itself and to build physically sound models for predicting the main objects and processes of solar activity.

\section{The global field components}

This work is mainly focused on the analysis of the large-scale global magnetic field. We represent the magnetic field as a sum of multipoles of different orders and study their time behavior separately. There is a large number of works devoted to the study of the behavior and evolution of the lowest degree multipoles (i.e., the largest scale), their amplitudes and phases, and their correlation with the solar photospheric magnetic field \citep{L77, H84, Getal92, GJ92, SV86, SW87, SG88, S94, KS05}. In these papers, the analysis is based on the Kitt Peak and WSO (John Wilcox Solar Observatory) data.

We decompose the surface field observed at the Wilcox Solar Observatory into its harmonic components and present the time evolution of the mode coefficients for the past three sunspot cycles. We have been working with the WSO (Stanford University) synoptic maps of the light-of-sight photospheric magnetic field component \citep{Setal77}   for the period from Carrington rotation 1642 (beginning on 27 May 1976) till rotation 2227 (beginning on 2 February 2020) converted into a sum of associated Legendre polynomials $P_l^m$; $g_{lm}$ and $h_{lm}$  are the Gauss coefficients calculated under the assumption that the magnetic field is potential between the photosphere and the source surface, while the magnetic field is presumed to be exactly radial at the source surface \citep{Oetal20}.

\begin{figure}
    \includegraphics[width=10cm]{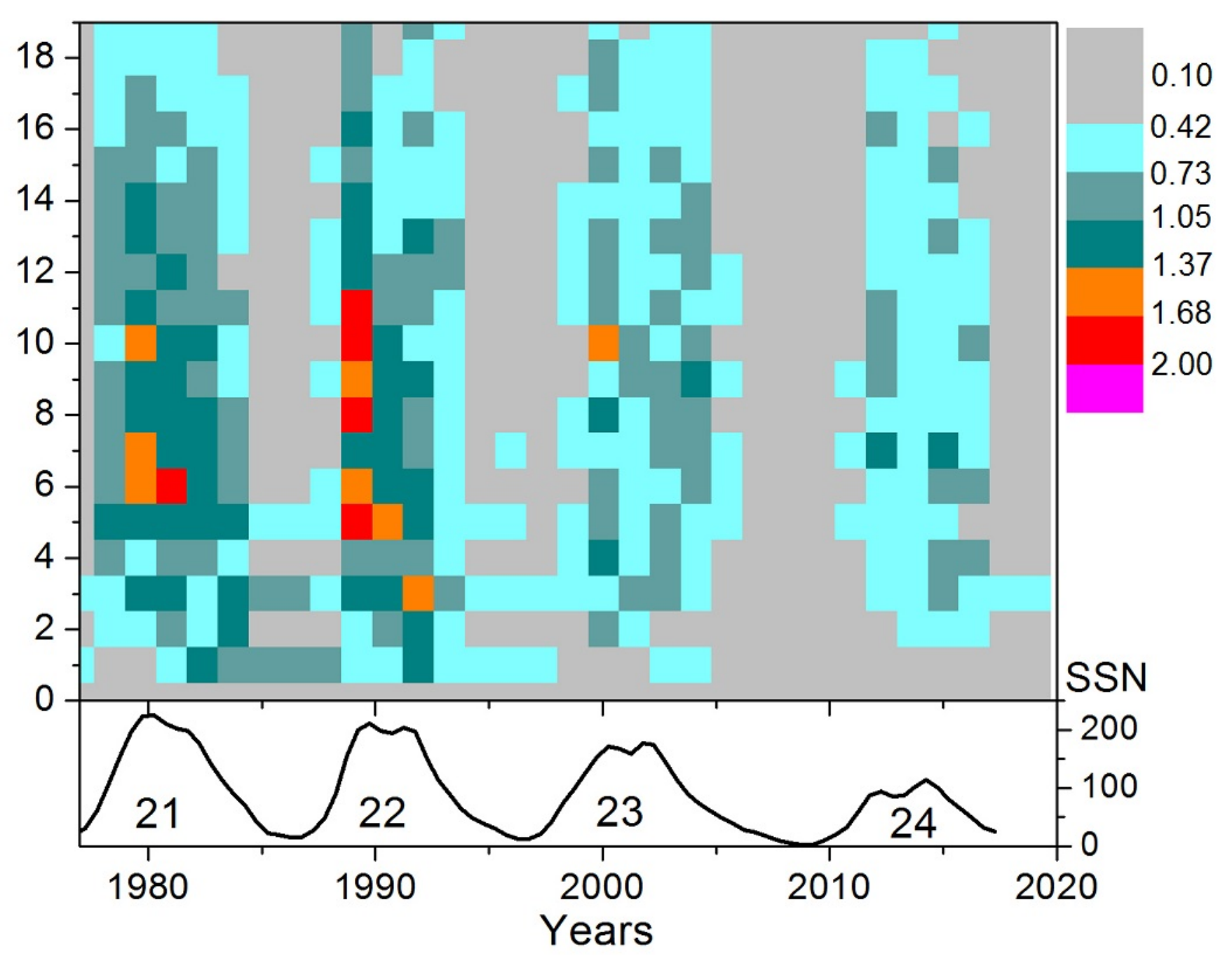}
    \caption{\label{fig1} Top panel - contribution of each of the first 20 harmonics ($l=0,1,2,\dots,19$) on the photosphere surface to the mean magnetic field versus time. The solar cycle according to the sunspot data is given for comparison at the bottom.} 
\end{figure}

Fig.~\ref{fig1} (top panel) represents the contribution of various l-harmonics to the mean surface magnetic field versus time. Some prevalence of the even harmonics is visible; however it is difficult to claim that the odd harmonics are unimportant. As seen from Fig.~\ref{fig1}, the contribution of the even and odd modes is comparable at the solar maximum, while at the solar minimum, the even modes become much smaller than the odd ones. All plotted harmonics become larger during the solar maximum, while the harmonics with $l = 1, 3, 5$ alone survive at the minimum. The contribution of harmonics with $l = 3$ and $l=5$ becomes most pronounced at the solar minimum. The even harmonics virtually disappear during the minimum. In the recent cycles, the time interval when the higher-order l-harmonics disappear has been getting longer; i.e. the poloidal field that forms the following cycle is lacking. However, in contrast to Cycle 23, after which all harmonics disappeared at the minimum, a quite significant third harmonic is visible at the minimum after Cycle 24.

\begin{figure}
    \centering
    \includegraphics[width=10cm]{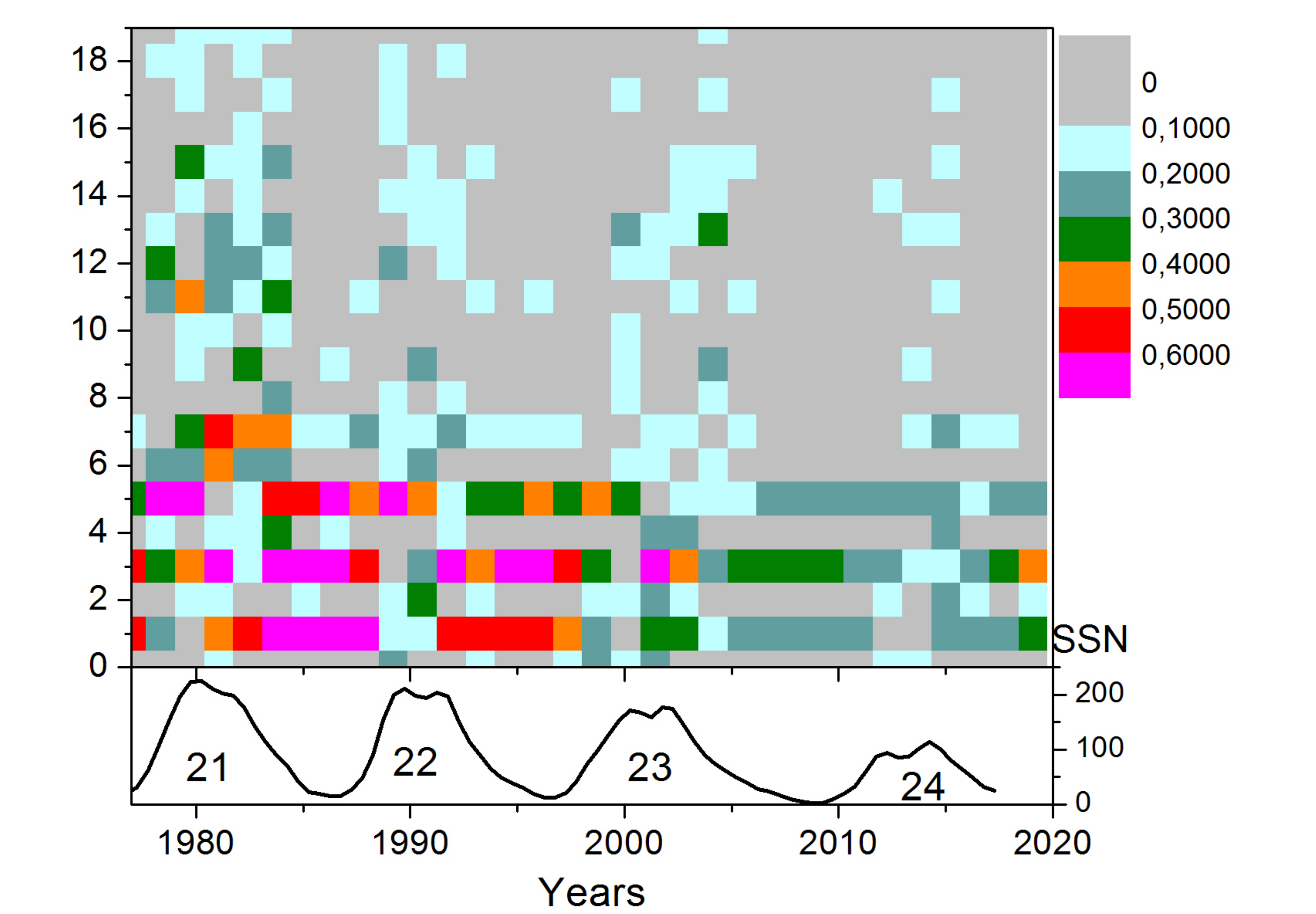}
    \caption{Contribution of each of the first 20 axisymmetric harmonics ($l = 0, 1, 2, \dots
19$) to the mean magnetic field at the photosphere surface versus time, $m = 0$ (top) The
black curve on the lower panel shows the solar cycle according to sunspot data.}
    \label{fig2}
\end{figure}

\begin{figure}
    \centering
    \includegraphics[width=10cm]{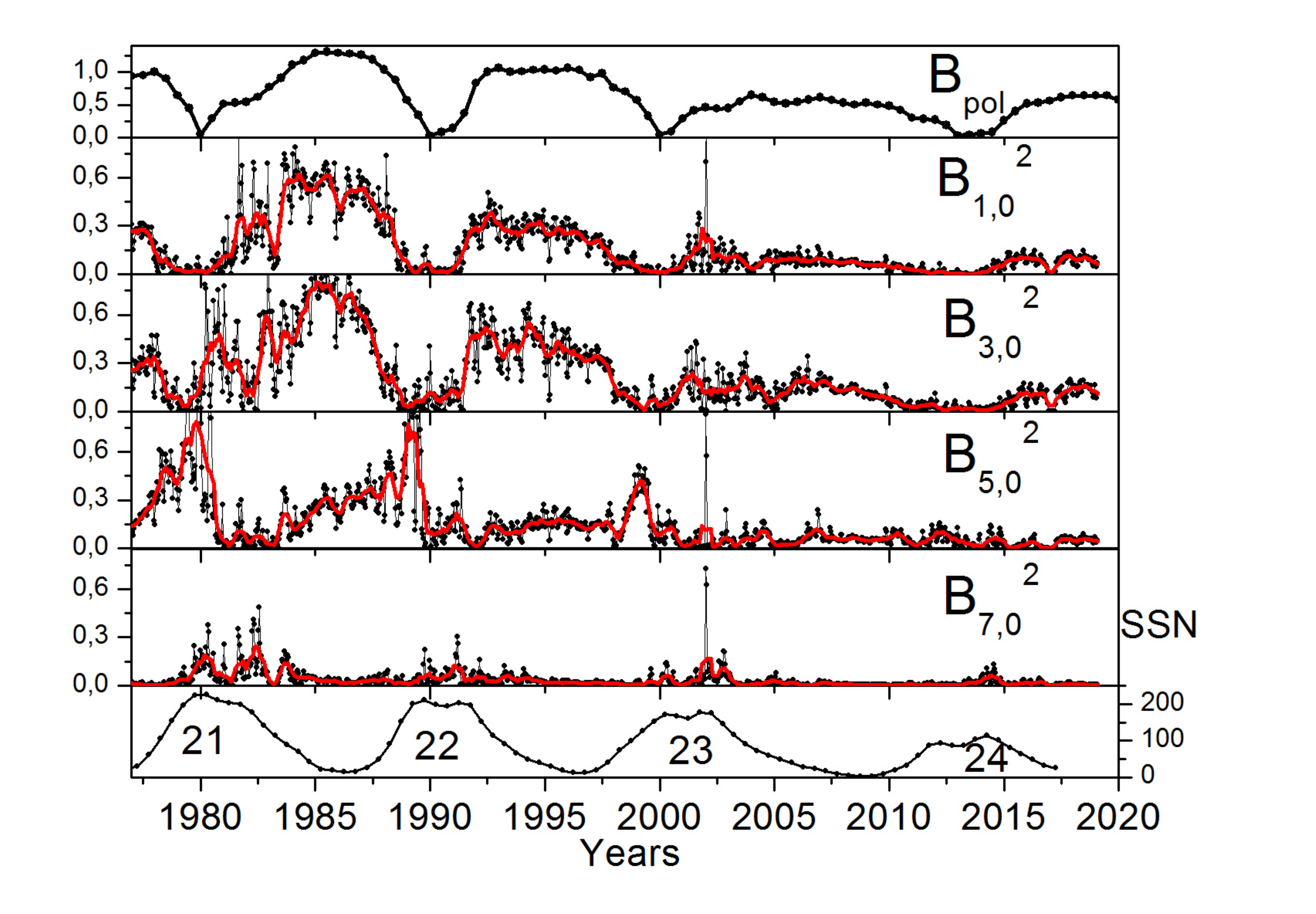}
    \caption{Time dependence of the first seven odd axisymmetric harmonics. The
lowest panel shows the time variation of sunspot numbers; the upper panel is the polar
field http://wso.stanford.edu/Polar.html).}
    \label{fig3}
\end{figure}

This effect is also present when we consider the axisymmetric harmonics alone (Fig.~\ref{fig2}). In this case, we can clearly see the prevalence of the 3rd harmonic in 2018-2020. 
Let us consider the cyclic variation of axisymmetric harmonics. This means that we shall analyze directly the harmonic coefficients $g_{l0}$.
Fig\ref{fig3} represents the first seven odd axisymmetric harmonics versus time. It is evident that harmonics 1, 3, and 5 are the strongest ones. This is exactly what \cite{S94} obtained from a different set of data. The seventh harmonic is the weakest. When passing to harmonics of higher order, one can notice a gradual shift of structures along the cycle, which was described by Stenflo. 

It is seen from Fig.~\ref{fig3} that only the first and the third harmonics are similar to the polar field in their behavior. The fifth and seventh harmonics are sharply different and resemble more closely the sunspot numbers. 

Let's note one more feature.  The axisymmetric harmonics $l = 1$ and $l=3$, $m = 0$ increase at the cycle minima and fall strongly at the maxima. However, the axisymmetric harmonic 5 behaves in a special way. It increases after the minimum in the growth phase. Also, the stochastic component for $l=5$ seems to be much stronger than that for $l = 1,3$.  In the non-smoothed time-series the harmonic $l=5$ tends to go in antiphase with harmonics 1 and 3.
Our previous analysis (see, \citealp{Oetal21}) showed that, for the smoothed 20-year-like oscillations, the phase difference between $l=5$  and $l = 1,3$ is about 2 years.
The results of the paper cited above show that the surface sunspot activity can produce a considerable effect on generation of the $l=5$ harmonic due to the emergence of the tilted bipolar active regions. It also can result in a seemingly antiphase evolution of $l=5$  and $l = 1,3$.

\begin{figure}
    \centering
    \includegraphics[width=10cm]{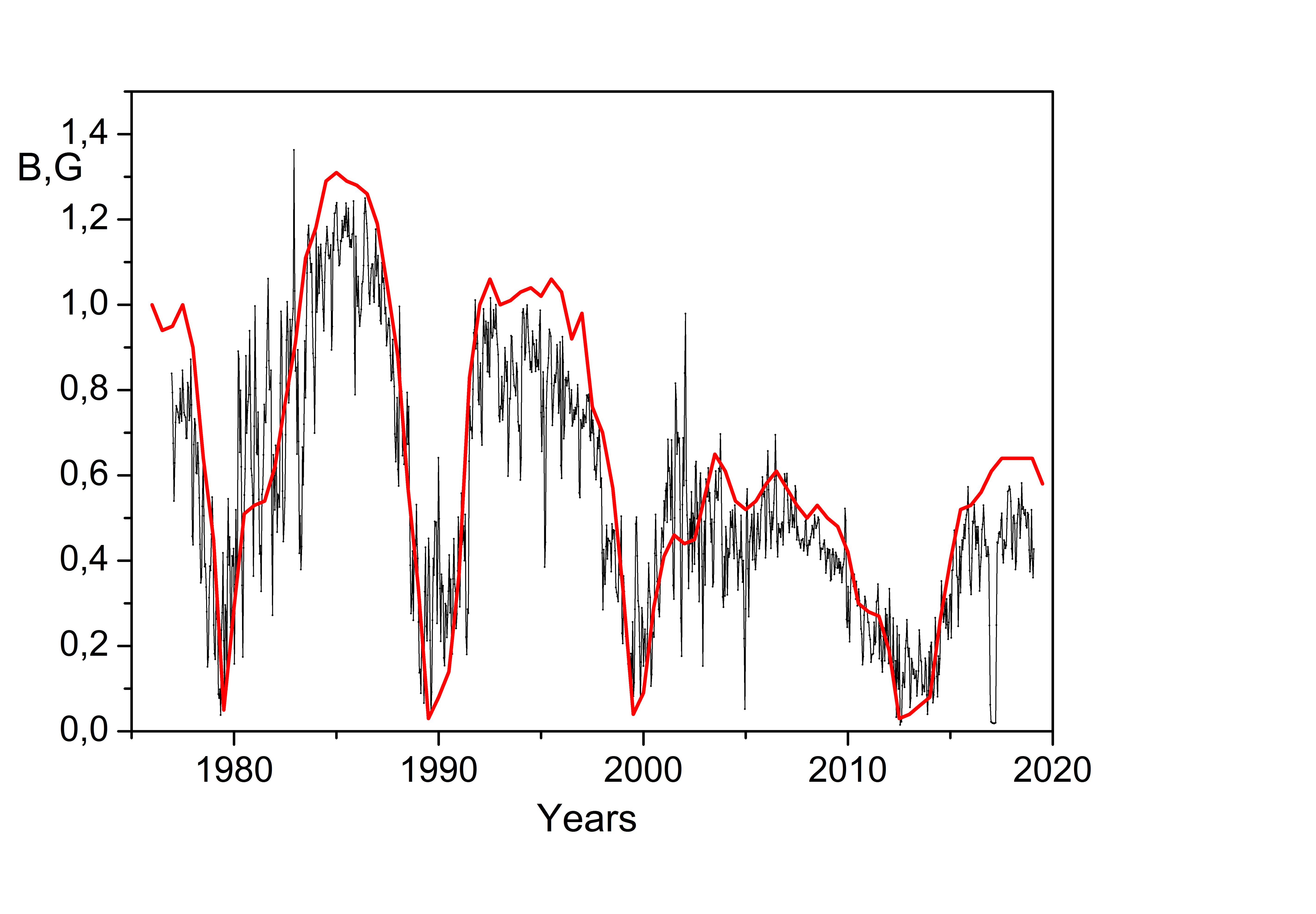}
\caption{The behaviour of the sum of the first and the third harmonics (black)  in comparison with the observed polar field (red). }
    \label{fig4}
\end{figure}

Moreover, the sum of the first and the third harmonics coincides quantitatively with the polar field proxy (see Fig.~\ref{fig4}). Despite some phase difference between the evolution of the polar magnetic field and $l = 1,3$ (see \citealp{Oetal21}), we see that the magnitude of the polar field is virtually determined by the largest scale, and it is apparently sufficient to find the first two harmonics to calculate it (also, cf. \citealp{Metal13}).

\section{Effective multipole index}

Now, following \cite{OE89} (see also \cite{OS92}),  we can calculate the mean square radial component of the magnetic field on the sphere of radius R.

\begin{equation}
i(B_r ) |_{R_0 }=\sum_{lm}\frac{(l+1+l \zeta^{2l+1})^2}{2l+1} (g_{lm}^2+h_{lm}^2 ),                                   \end{equation} 
\begin{equation}
i(R_r ) |_{R_s}=\sum_{lm}(2l+1) \zeta^{2l+4} (g_lm^2+h_lm^2 ),                                      
\end{equation}
where $\zeta=R_0/R_s$. 
This means that the contribution of the l-th mode to the mean magnetic field contains an l-dependent coefficient. In these formulas, $i(B_r ) |_{R_0}$ and $i(R_r ) |_{R_s}$ are the mean square radial components of the magnetic field in the photosphere and at the source surface, respectively. In our calculations, the radius of the source surface $R_S =2.5 R_0$ and, thus, $\zeta=0.4$.

\begin{figure}
    \centering
    \includegraphics[width=10cm]{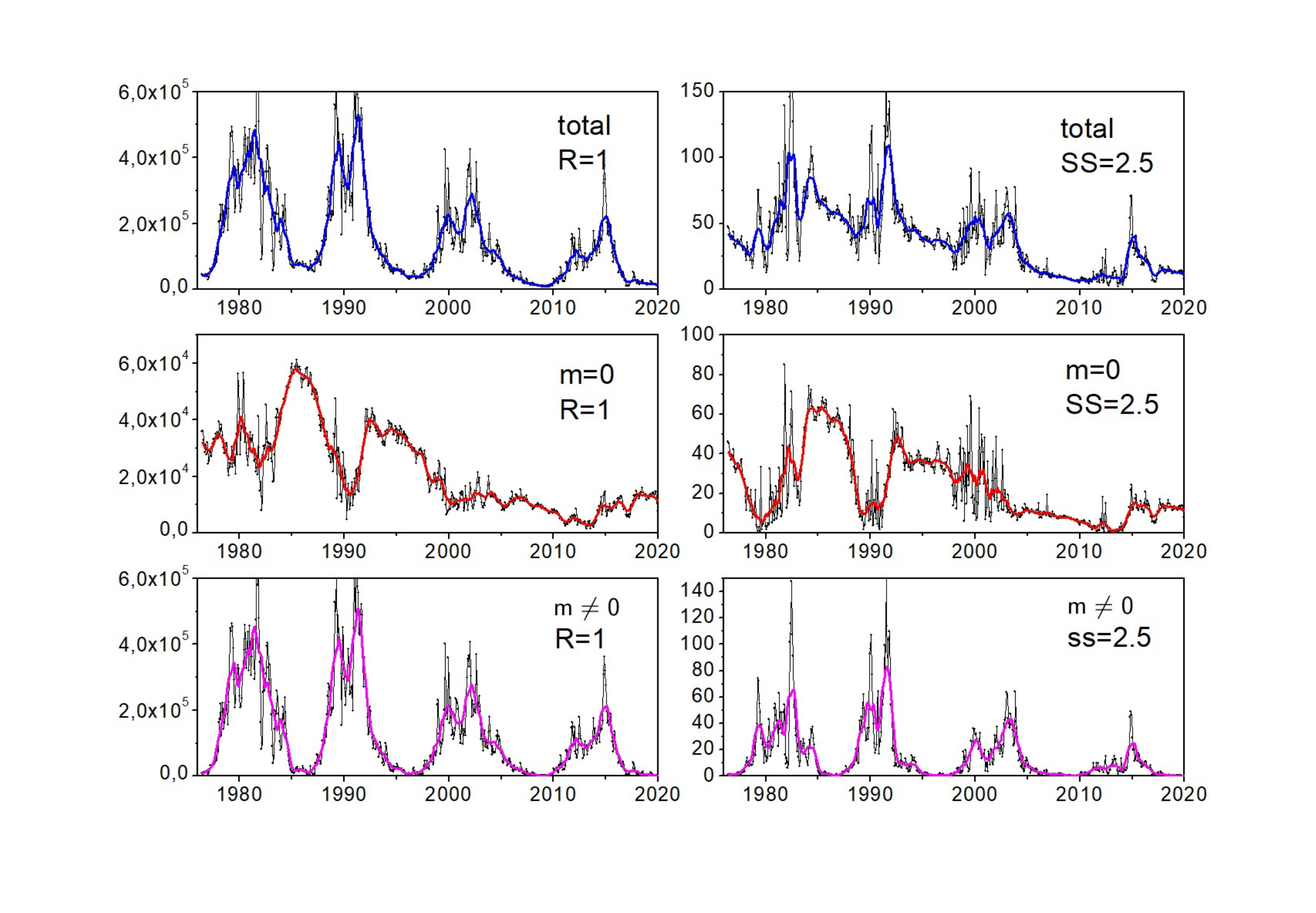}
\caption{$i(B_r)$ index for the photosphere (left panel) and for the source surface (right
panel). All values are in microtesla squared. The black curves are drawn in every half
the Carringoton rotation; the color curves represent smoothing over 13 points.  }
    \label{fig5}
\end{figure}

We can calculate these indices separately for the axisymmetric or nonaxissymmetric components; i.e., put $ m = 0$ or $m \ne 0$ in Eqs. (1) - (2). These curves are shown in Fig.~\ref{fig5}.

Fig.~\ref{fig5} shows the i(Br) index for the photosphere and for the source surface. If we compare the two plots, we will also see here an 11-year periodicity; the phases of the maxima mainly coincide; though in the large-scale field, one can more clearly see the Gnevyshev dip  \citep{S79} at the cycle maximum. As a result, the secondary maximum is more clearly pronounced, too. The maxima of the global magnetic field are shifted by 2-3 years relative to the sunspot maxima. 
One can also see other features such as a gradually decreasing activity of the fields of all scales. It is this decrease that makes us believe that we are on the eve of a very low cycle. At first glance, the decrease is approximately the same in the photosphere and at the source surface. However, this is not exactly so.
At the same time, the axisymmetric component m=0 behaves completely differently. First of all, there is a much more pronounced decline in the magnitude of the heights to later cycles. The maxima themselves are shifted relative to the maxima of the total field. Moreover, Cycle 23 (1997-2008) is almost not noticeable at all. This is a direct consequence of the difference in the behavior of harmonics 1-3 and 5 (see Figure 3).

To assess the contribution of various components, we can use the effective multipole index introduced in our earlier work.

\begin{equation}
I_M=-\frac{1}{2\log 2.5} \log\frac{i(B_r)|_{R_s}}{i(B_r)|_{R_0}}.         
\end{equation}
It is determined as the logarithm of the ratio of $i(R_r )$ at the source surface to its value in the photosphere \citep{Ietal97}. In fact, when passing from the photosphere to the source surface, the mean square radial component of the magnetic field changes in accordance with the power function $\sim R^{2n}$, i.e., $\sim 2.5{-2n}$ , where $n = 3$ for a dipole source, 
$n = 4$ for a quadrupole source, and $n > 4$ for a higher-order multipole source. When the field under discussion is a combination of fields from several sources with different weight, $n$ can assume values from 3 to 4 (in the case of combined dipole and quadrupole sources) or higher (in the case of a higher-order multipole field).

\begin{figure}
    \centering
    \includegraphics[width=10cm]{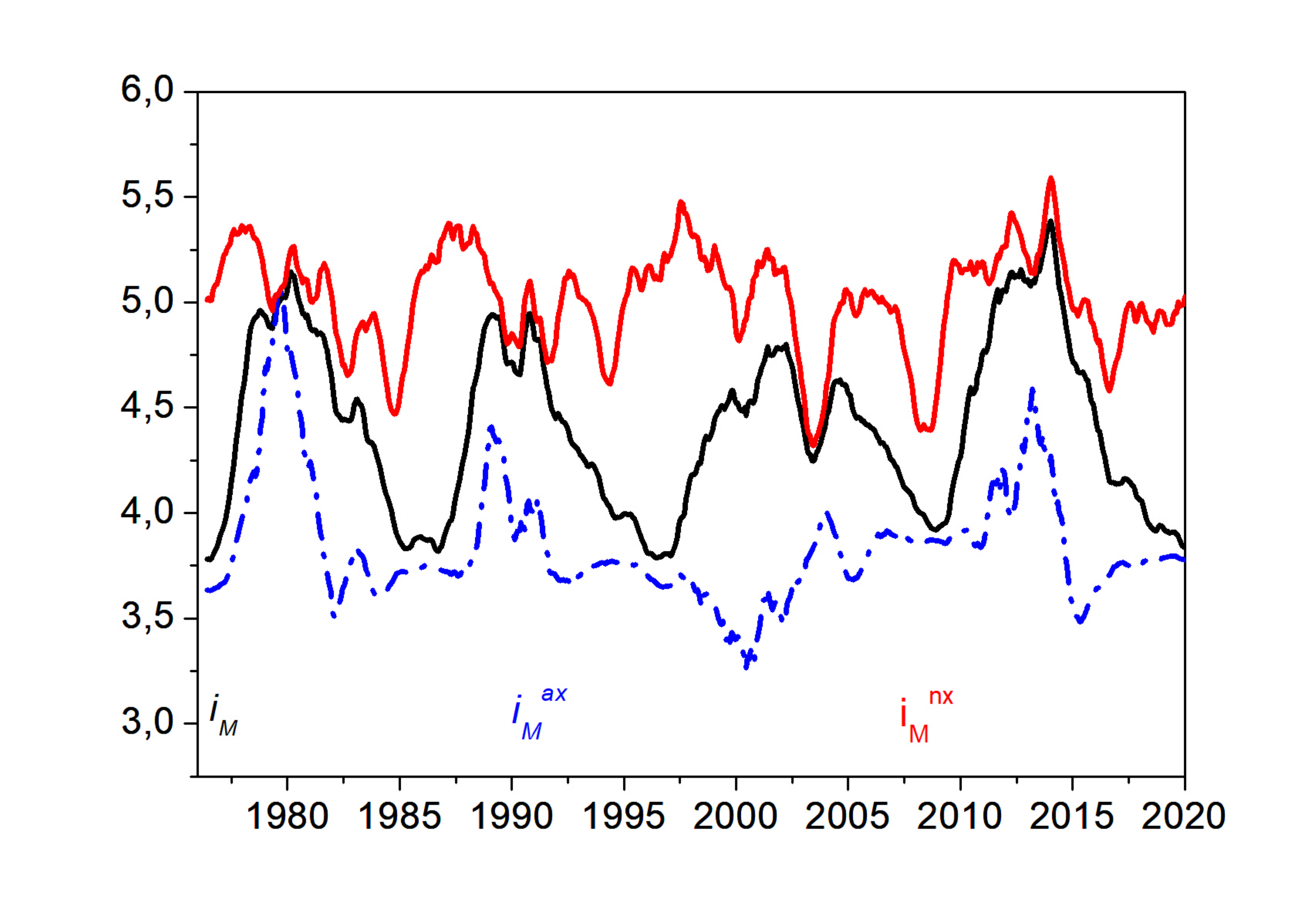}
\caption{Index of the effective multipole (black curve - total, blue curve - the axisymmetric components alone, m=0, red line - nonaxisymetric).  }
    \label{fig6}
\end{figure}

We see from Fig.~\ref{fig6} that Cycle 24 was abnormal as concerns the gradual decrease of the effective multipole index. The $I_M$ index increased sharply, which could only be the result of a more rapid decrease of the field with height. This, in turn, should imply an increased contribution of higher-order harmonics and, accordingly, a transition to components of smaller scale. The index of the non-axisymmetric field, as expected, is mainly concentrated near the values of $5.5 \dots  5$, falling to lower values during the periods of the cycle minima and during the periods of Gnevyshev dips in the phases of the maximum. At the same time, the axisymmetric field index behaves quite unexpectedly; it reaches a very low value at the beginning of the maximum of Cycle 23 and then, rises slightly at the end of the maximum in 2004.

\section{Index of effective magnetic multipole and other solar activity indices}

The solar activity calendar is based on standard sunspot numbers. They underlie the 11-year cycle of solar activity. This 11-year cycle with certain phase shifts is revealed in the majority of the other indices, as well. However, there are significant differences as concerns the amplitude of the extreme values of individual cycles.

 During the past four cycles, the heights of the sunspot maxima were successively decreasing. However, this pattern is not quite true for the magnetic indices. E.g., the polar field at the minimum after Cycle 24 is somewhat larger than it was at the minimum after Cycle 23 (see Figs. 3 and 4). Cycle 23 is weakly pronounced in the axisymmetric field, at least much weaker than the adjacent Cycles 22 and 24. The particular features in the behavior of the axisymmetric field may be due to the fact that, as mentioned above (see Fig. 3), the fifth harmonic is in antiphase with the first and third ones. This particularity is especially evident in Cycle 23. And finally, the index of the effective magnetic multipole is significantly higher in Cycle 24 than in the previous Cycle 23.
It was also found that the magnetic field in the sunspot umbra decreased significantly (down to ~2050 G; i.e., by ~700 G relative to the level of 1998) \citep{Letal12, LW15} (Livingston, Penn, and Svalgaard., 2012; Livingston and Watson, 2015), probably, due to a significant increase in the relative number of small sunspot groups \citep{Netal12}.

\begin{figure}
    \centering
    \includegraphics[width=10cm]{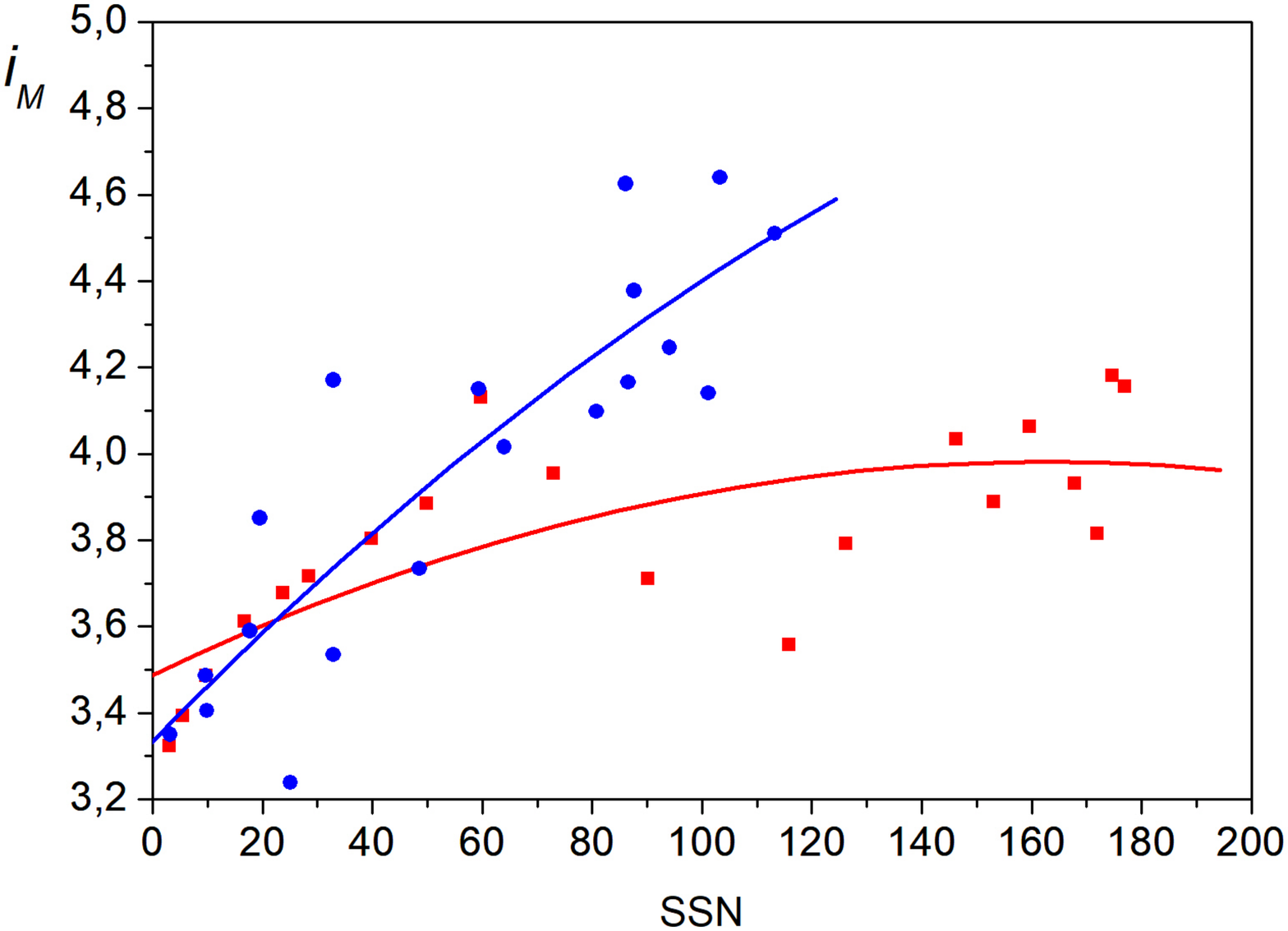}
\caption{The $i_M$ index as a function of the sunspot number SSN. The values for
Cycle 23 are shown in red; the values for Cycle 24 are shown in blue. The red and blue lines are the
corresponding approximation by a second-order polynomial. }
    \label{fig7}
\end{figure}

Fig.~\ref{fig7} shows the dependence of the iM index on the sunspot number SSN separately for Cycles 23 and 24. One can see that the ratio of the magnetic scales in these two cycles is completely different. In Cycle 23, despite a higher level of SSN, the transition to higher-order harmonics is slowed down. In Cycle 24, active features of smaller scales appear much earlier and exist during the entire cycle. I.e., the relationship between the large-scale magnetic field ($i_M$ index) and local fields (SSN index) is ambiguous and changes over time.

\begin{figure}
    \centering
    \includegraphics[width=10cm]{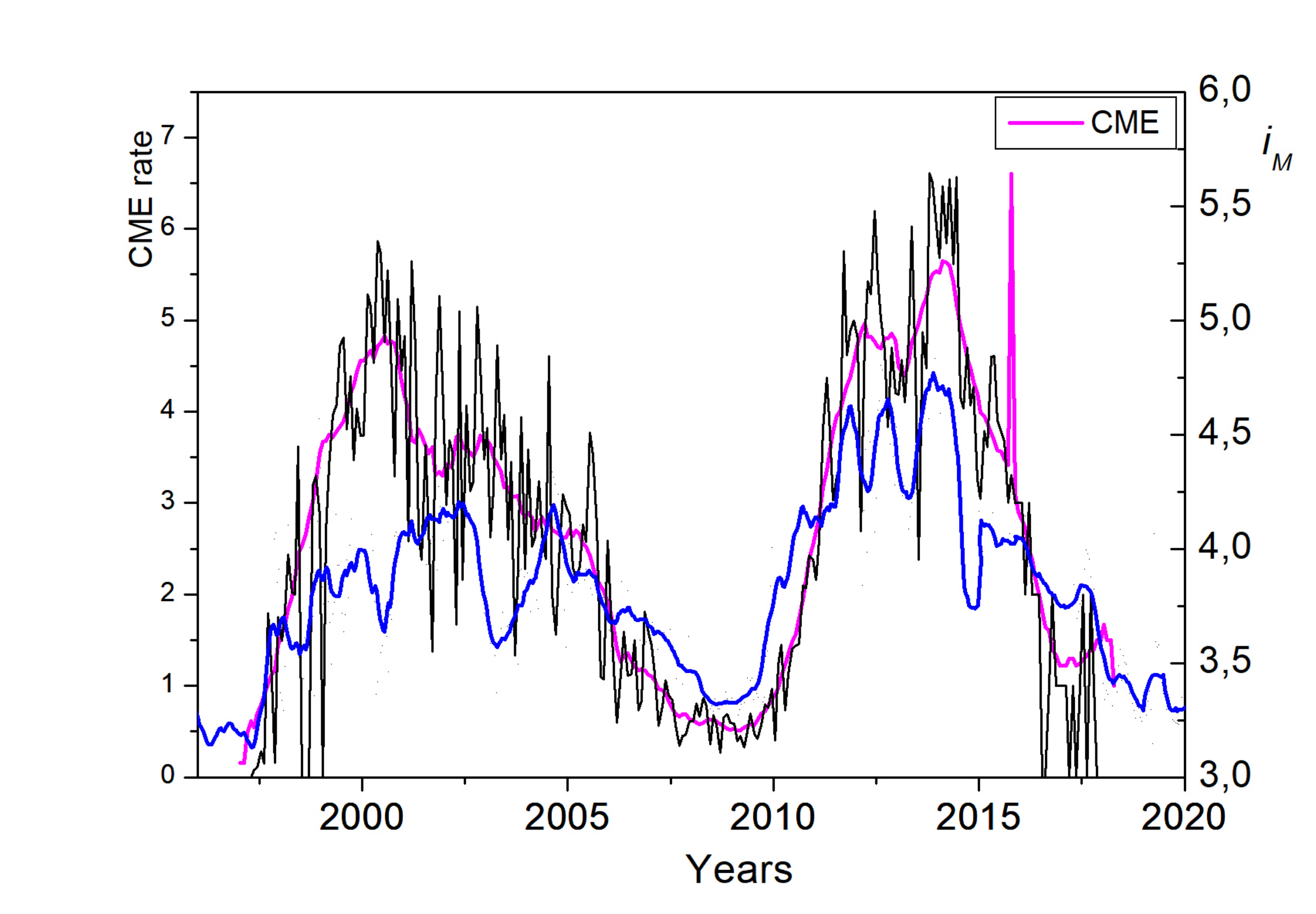}
\caption{Time variation of the monthly mean CME occurrence rate (black); the same smoothed over 13 months (magenta); and $i_m$ time variation smoothed over 13 months (blue).}
    \label{fig8}
\end{figure}

It should be noted that despite the fact that Cycle 24 was extremely low, the number of coronal mass ejections recorded at that time was much larger than in Cycle 23 ($http://sidc.oma.be/cactus$). Fig.~\ref{fig8}  compares the  CME occurrence rate with $i_M$. 

\begin{figure}
\includegraphics[width=1.\textwidth]{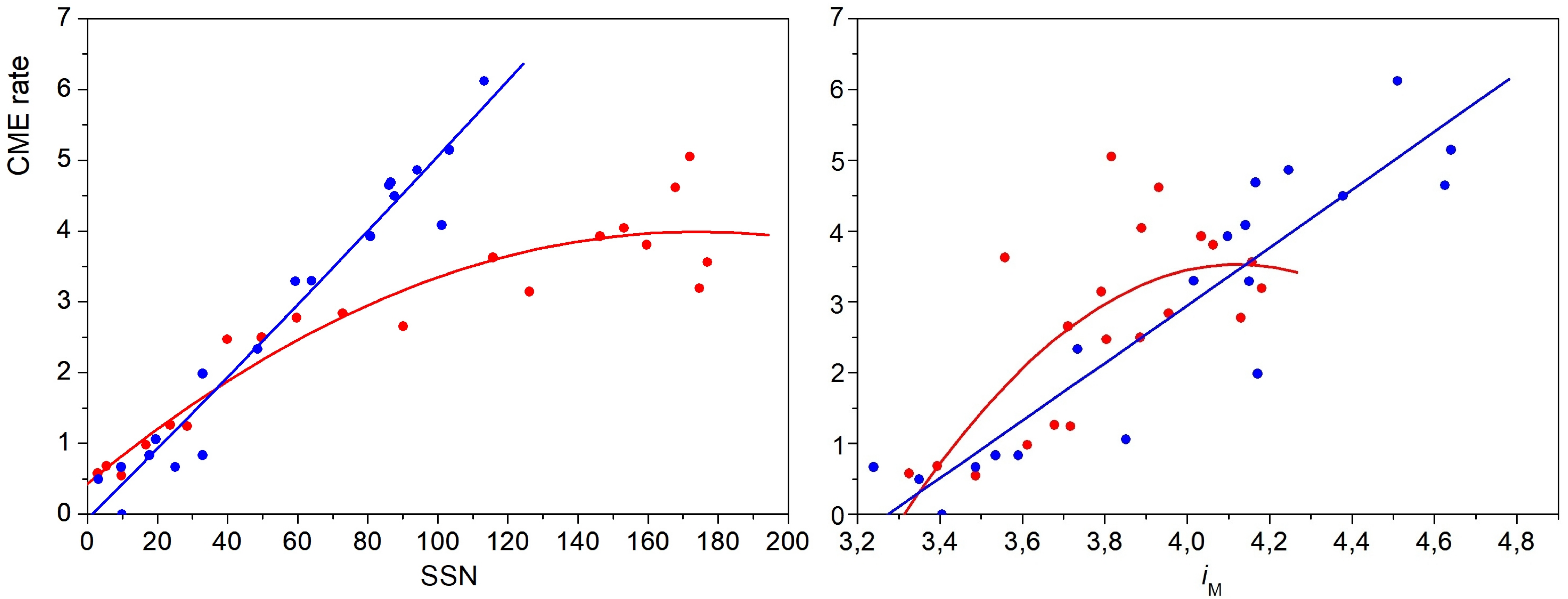}
\caption{CME monthly numbers  (left) and $i_M$ (right) versus SNN; red stands for Cycle 23 and blue stands for Cycle 24. \label{fig9}}
\end{figure}

{Fig.~\ref{fig9} shows the link between CME and SSN (left) and iM (right) separately
for Cycles 23 (red) and 24 (blue). Figures 7, 8 and 9 show that the relationship between the number of sunspots,  the effective multipole  index and the number of coronal mass ejections in cycle 24 has changed dramatically. The number of spots has decreased, while the other two indices have increased. But the dependence of the CME indices on SSN and iM within the cycles is approximately the same. Therefore, the CME daily occurrence rate can be determined not only by widely discussed link with the sunspot number but also by the total magnetic
field structure measured by the effective multipole index.}

A similar effect was revealed by \cite{CB17}. They showed that the fraction of SDE $\ge$ M1.0-class flares (including spikes) in the weaker Cycle 24 exceeded that observed in Cycle 23 for all three temporal parameters in the maximum phase of the cycle and for the decay parameter, in the raise phase. They also arrived at a conclusion that the fraction of SDEs turned out to be markedly increased at the beginning of Cycle 24 perhaps due to the fact that the proportion of small spots also increased in 2010-2011.

These two effects may be the result of an increased relative role of the higher-order harmonics in Cycle 24.

\begin{figure}
    \centering
    \includegraphics[width=10cm]{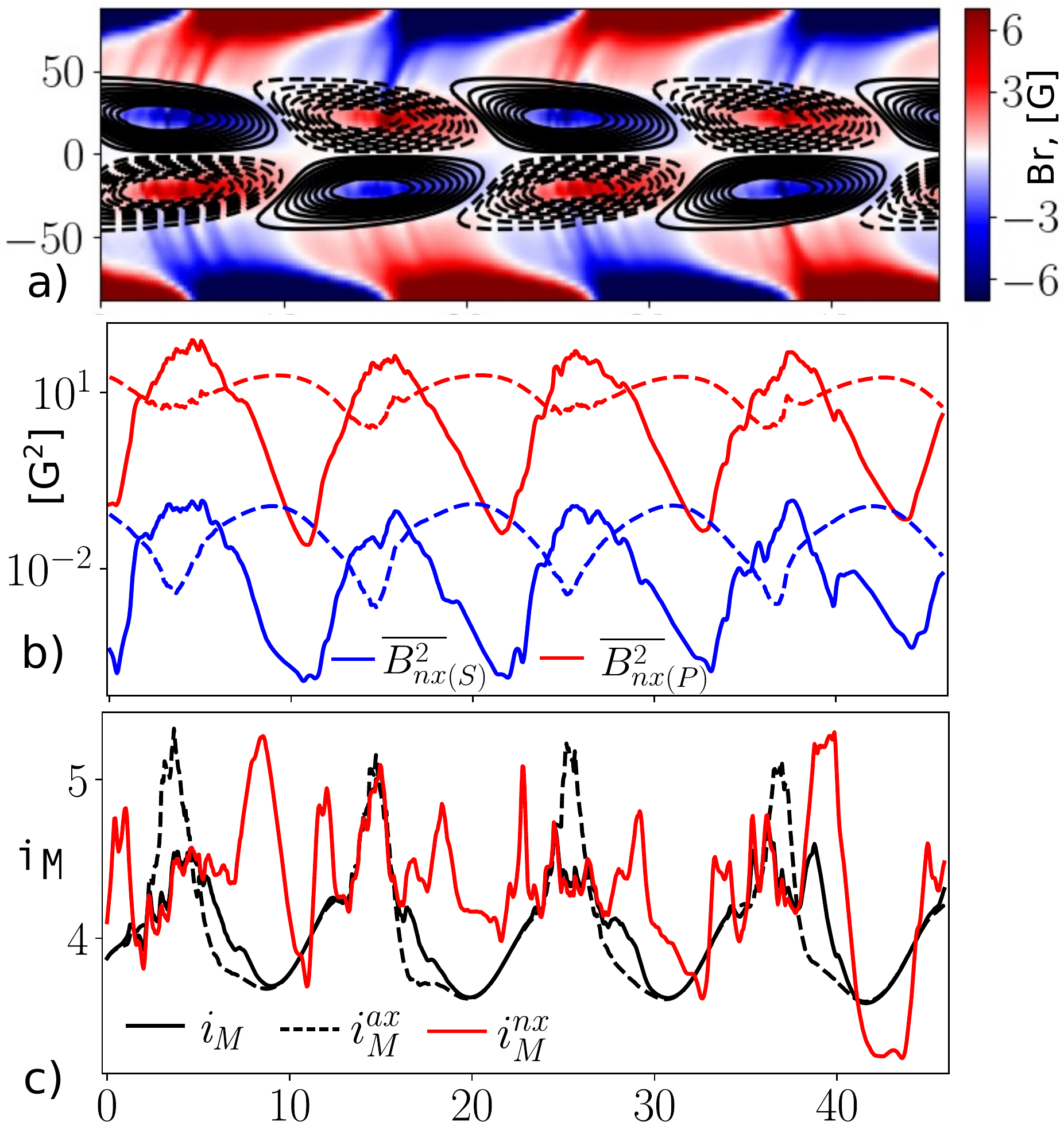}
\caption{Cyclic variation of the index of the effective magnetic multipole based on the dynamo model by \protect{\cite{Pipin2021} and \cite{Oetal21} }. { Panel (a) shows the time-latitude diagrams of the surface radial magnetic field (background color) and the toroidal magnetic field in the subsurface shear layer (levels are in range of $\pm 1$kG; b) the solid line shows the mean-square density of the radial magnetic field at the surface, the dashed red line shows the same for the axisymmetric part of the magnetic field and  the blue lines show the same for the source surface level; c) the  black solid line shows the multipole index evolution, the dashed line and the red line show the same index computed from the axisymmetric and nonaxisymmetric components of the magnetic field, respectively.}}
\label{fig10}
\end{figure}

The cyclic variation of the effective magnetic multipole index can be simulated using the dynamo model by \cite{Pipin2021} and \cite{Oetal21}. The results of this calculation are shown in Fig. 10. The agreement of these results with observations is beyond doubt. As well as in the observation results, one can see a cyclic variation of all components. The main difference is that the model does not include cycle-to-cycle variations, in particular, the behavior of the axisymmetric component.

\section{Discussion and Conclusions}

%\section{A message for solar activity forecast}

We see in some figures presented in this paper that the large-scale poloidal magnetic field gradually declines from cycle to cycle to reach very low values in the past cycle. The decline is most pronounced in the lowest-order modes. According to the scheme of action of the solar dynamo, the poloidal magnetic field and the corresponding lowest-order modes, in particular, can be considered a source of production of the toroidal magnetic field by the solar dynamo engine. If we correctly assess the role of the drivers of cyclic solar activity, it is natural to expect that the next cycle will be weak as recorded in sunspot data. On the other hand, the behavior of the polar field in the past two cycles infers that the next solar activity cycle as recorded in sunspot data will not be lower than the previous one. Indeed, the polar field near 2020 is slightly larger than it was near the previous solar maximum. As seen above, the dynamics of the near-surface magnetic field contributes substantially to the dynamo action; so, the surface flux-transport mechanism is indeed important to the solar dynamo. In other words, messages from the polar magnetic field and from the lower modes of the large-scale surface magnetic field do not fully coincide, which opens up an opportunity to clarify details of regeneration of the poloidal magnetic field from the toroidal one. It should be noted here that the polar field is not by far the most reliable predictor of the following cycle. \cite{Metal13} showed that the correlation coefficient between the polar field and the height of the following cycle is only 0.60.

Our results suggest that the global minimum of solar activity both in the local and in the large-scale fields may have already been reached in Cycles 23-24. This conclusion does not only follow from the analysis of the polar field, but it was also refined by analyzing the zonal harmonics. The zonal harmonics are measured and calculated more accurately than the directly measured polar field. The point is that the components of the magnetic field vector measured with a magnetograph near the limb are not the same as those measured in the central part of the solar disk. Therefore, restoring the polar field from the sum of the first and the third zonal harmonics has its advantages. In general, it can be expected that Cycle 25 will continue the series of low cycles comparable in height to Cycles 23 and 24.

{Note also that using the data on the fifth harmonic  it is possible to specify the date of the cycle maximum in  advance. Figure 3 shows that the fifth harmonic revives from a deep minimum earlier than other global harmonics, 1-2 years before the SSN maximum and before the date of the polar field reversal. This is possibly due to the fact that early manifestations of the sunspot cycle are observed at mid-latitudes and are associated with relatively small scales.}	

Let us summarize again the results of our paper. We suggest that there are parameters of the  zonal harmonics of the solar surface magnetic field, such as the magnitude of the $l=3$ harmonic or the effective multipole index, that can be used as a reasonable addition to the polar magnetic field proxies.

Note that we do not consider here the particular methods of prediction of solar cyclic activity because apart from the  physical mechanisms selected for the forecast, such discussion should include various other topics like the aims of a particular forecast, verification of the forecast within the framework of observational data as well as dynamo modelling, which is far out of the scope of this paper.

Of course, we fully appreciate a possible role of other tracers that can possibly support the polar field data in the problem of the solar activity forecast, in particular, the geomagnetic index (the idea originated in \cite{OO79} was exploited in the recent paper \cite{Betal21}).

\textbf{Acknowledgements}
The authors are grateful to the WSO teams for free access to their
data. V. Obridko, D. Sokoloff , and A. Shibalova thank the support by
RFBR, Russia grants nos. 20-02-00150, and 19-52-53045. In addition,
DDS and AAS thanks support by BASIS, Russia fund numbers 18-1-1-
77-3 and 21-1-1-4-1. All the authors thank the financial support of the
Ministry of Science and Higher Education of the Russian Federation:
VVP thanks agreement no. 075-GZ/C3569/278; VNO, DDS, AAS and IML
thank agreements No. 075-15-2019-1621, and 075-15-2020-780.

%\appendix
%\section{My Appendix}

%\verb+\printcredits+ command is used after appendix sections to list 
%author credit taxonomy contribution roles tagged using \verb+\credit+ 
%in frontmatter.

%\section*{References}
\bibliographystyle{elsarticle-harv}

%\bibliography{dyn}
%\input{obridkoetal.bbl}


\end{document}